%% file: paper.tex
\documentclass[pre,twocolumn,showpacs,preprintnumbers,superscriptaddress]{revtex4}
\usepackage{graphicx,bm,times,url}
\usepackage{amsmath}
\graphicspath{{./fig/}{./png/}}
\newcommand{\eal}{{\itshape{et al.}}}

\input{symdef.inc}

\begin{document}
\title{Breakdown of chiral symmetry during saturation of the Tayler instability}
\author{Alfio Bonanno}
\affiliation{INAF, Osservatorio Astrofisico di Catania, Via S. Sofia 78, 95123 Catania, Italy}
\affiliation{INFN, Sezione di Catania, Via S. Sofia 72, 95123 Catania, Italy}
\author{Axel Brandenburg}
\affiliation{Nordita, Royal Institute of Technology and Stockholm University,
Roslagstullsbacken 23, SE-10691 Stockholm, Sweden}
\affiliation{Department of Astronomy, Stockholm University, 
     SE 10691 Stockholm, Sweden}
\author{Fabio Del Sordo}
\affiliation{Nordita, Royal Institute of Technology and Stockholm University,
Roslagstullsbacken 23, SE-10691 Stockholm, Sweden}
\affiliation{Department of Astronomy,
Stockholm University, SE 10691 Stockholm, Sweden}
\author{Dhrubaditya Mitra}
\affiliation{Nordita, Royal Institute of Technology and Stockholm University,
Roslagstullsbacken 23, SE-10691 Stockholm, Sweden}

\date{\today,~ $ $Revision: 1.191 $ $}

\begin{abstract}
We study spontaneous breakdown of chiral symmetry during the nonlinear
evolution of the Tayler instability.
We start with an initial  steady state of zero helicity.
Within linearized perturbation calculations, helical perturbations of
this initial state have the same growth rate for either sign of helicity. 
Direct numerical simulations (DNS) of the fully nonlinear equations,
however, show that an infinitesimal excess of one sign of helicity in the initial
perturbation gives rise to a saturated helical state. 
We further show that this symmetry breaking can be described by 
weakly nonlinear finite--amplitude equations with undetermined
coefficients which can be deduced solely from symmetry consideration. 
By fitting solutions of the amplitude equations to data from DNS
we further determine the coefficients of the amplitude equations. 
\end{abstract}
\maketitle
\section{Introduction}

There are many examples in nature where the ground state does not
share the same symmetries of the underlying
equations of motion \cite{ume}.
The most common examples 
include equilibrium phase transition, e.g., 
the case of a liquid-solid transition where the space translational
symmetry is broken, or that of a paramagnetic-ferromagnetic
transition 
where the spin-rotational symmetry is broken; see e.g., Ref.\cite{Gol}
for a detailed discussion. 
However, the original symmetry is not lost but gives rise to the appearance of
a regular structure with a specific length scale.

In non-equilibrium physics spontaneous symmetry breaking is often
observed when some control parameter is increased above a critical
value, see e.g., Ref~\cite{rb} for a comprehensive introduction.  
Two well--studied examples from fluid dynamics include the case of Rayleigh--B\'enard convection \cite{rb} 
and the Mullins--Sekerka instability of a moving interface between two phases \cite{muse}.   
These systems, too, are invariant under translation and reflection, but the basic instability 
produces a symmetry breaking bifurcation in which the continuous translational symmetry of the basic 
state is broken to a 
discrete one, although the mirror symmetry is often retained.
If the instability parameter is raised further, secondary instabilities may 
break the periodic pattern and eventually a completely new
symmetry-broken state may emerge, as has been 
seen in several experiments \cite{exp}.
At very high values of the control parameter,
turbulence sets in and most of the symmetries are
statistically restored.

In a hydrodynamic system under rotation, spontaneous
breakdown of chiral symmetry has been studied; see, e.g., Ref.~\cite{pinter06}.  
Spontaneous chiral symmetry breaking is also found
in liquid crystals \cite{seli93}. 
Preliminary evidence showing spontaneous chiral symmetry 
breaking in magnetohydrodynamics (MHD), in the absence of 
rotation, has been presented for the magnetic buoyancy
instability \cite{cha+mit+bra+rhe11} 
and for the Tayler instability in a Taylor-Couette setup \cite{gel+rud+hol11}.
However the role of the dynamics of the 
bifurcation process is still poorly understood.

The purpose of the present paper is twofold.
First we demonstrate the occurrence of spontaneous chiral symmetry breaking
in the context of a global instability of the toroidal field,
and second we elucidate some aspects
of the underlying nonlinear mechanism which determines the evolution
from a mirror-symmetric state to a state with a preferred handedness
or helicity.
In particular, we shall be interested in the case of the Tayler instability
\cite{tay73a,tay73b}, which has attracted much interest in recent
times for its possible astrophysical applications \cite{bo08a,bo08b, bu11,bu12, brano06,bra06, gel+rud+hol11,spru99}.
We thus discuss the possibility of generating a final state with finite
helicity starting from a non-helical basic
state, using a very small controlled helical perturbation.

Our setup has the advantage of better clarifying the complex nonlinear coupling between the different modes, which 
eventually leads to the formation of a final helical state.
In fact, the Tayler instability, in its simplest realization, has no
threshold field, at least
in ideal MHD \cite{bo08b}, where a sufficient condition for instability simply reads 
\begin{equation}\label{eq:stability}
\beta \equiv \frac {\partial \ln B_\varphi}{\partial \ln s}  > -\frac{1}{2},
\end{equation}
$s$ being the cylindrical radius. 
On the other hand, the spectrum is characterized by an infinite number
of unstable modes all characterized by pairs of opposite
azimuthal wave number $m=\pm 1$, 2, 3, ...., but with precisely the same
growth rate. In particular, as is well known, 
$m=\pm 1$ are the modes with the fastest growth rate.
Here our aim is to understand the dynamics of the bifurcation
process which leads to the selection of a final state of finite helicity
and to understand the evolution of the system after the bifurcation
takes place. 
It should also be noted that in the linear stage the preferred helicity is determined essentially by the 
helicity of the perturbation, but the nonlinear evolution can be rather complex and it is not clear
a priori what the final selected helical state would be.

The rest of the paper is organized as follows. In
Section~\ref{Amplitude} we write down the 
finite-amplitude equations that govern the evolution of the
instability in the weakly nonlinear phase. 
Our approach is based on symmetry arguments; a detailed analytical
derivation of the amplitude equations is avoided here. 
We find that the amplitude equations predict a breakdown of parity
for certain choice of parameters. 
Direct numerical simulations (DNS) of the fully nonlinear equations
describing the evolution of the Tayler instability are performed in
Section~\ref{DNS}. 
In our DNS studies we also find breakdown of parity. 
We fit data from DNS to solutions of the amplitude equations to
numerically determine the parameters appearing in the amplitude
equations.
It turns out that the amplitude equations we deduce are identical
to those used to describe the breakdown of mirror symmetry
in studies of the biochemical origin of life.
This connection is explored in
Section~\ref{Homochirality}. Finally conclusions are drawn in Section~\ref{Conclusion}

\section{Amplitude equations}
\label{Amplitude}

To the best of our knowledge
the amplitude equations describing the spontaneous breakdown of mirror
symmetry in hydrodynamic instabilities were first described in Ref.~\cite{fau+dou+thu91}.
The basic idea is as follows. 

Let us consider an instability with two growing modes with opposite helicity
but exactly the same growth rate and
let the amplitude in this basis of the left- and right-handed modes be given by
vectors $\hatL$ and $\hatR$, respectively. 
In physical space we have
\begin{eqnarray}
 \L (\bx) &=&  \hatL \phi(\nn), \label{eq:L}\\
 \R (\bx) &=&  \hatR \phi(\nn). \label{eq:R}
\end{eqnarray} 
For example, in Cartesian domains, with real--space coordinate $\xx$,
 $\phi(\boldmath$n$) = \exp(i\nn\cdot\bx)$. 
In cylindrical coordinate, $\phi$ is a combination of trigonometric and
Bessel functions. 
As the modes are helical they satisfy the Beltrami relation, 
\begin{equation}
\curl \R = \Lambda \R \quad\mbox{and}\quad \curl \L = -\Lambda \L .
\end{equation}

For the present problem, explicit expressions involve a linear combinations of
the type $J_m( s\sqrt{\Lambda^2+n^2 \pi^2/h^2}) \cos(m\phi) \cos( z n\pi /h)$,
where $J_m$ is the Bessel function of the first kind, $n,m=\pm1,2,3\dots$,
$h$ is the height of the cylinder and $s$ the cylindrical radius \cite{qing}. 
The set of such modes forms
a complete set (a Hilbert basis)
 for the spatial distribution of the field.

Here we assume that the dynamical evolution of the unstable mode
is determined by an effective Lagrangian.
For the left-handed helical mode, total helicity and energy are given by
\begin{eqnarray}
\El &=&\frac{1}{2} \int \L^2 (\bx) \, d^3 x
= \frac{1}{2}  \hatL \cdot \hatL^{\ast}, \label{eq:EL} \\
\cHl &=& \int \L \cdot \curl \L \, d^3 x = - 2\Lambda \El.
\label{eq:eandh}
\end{eqnarray}
where an asterisk denotes complex conjugation. 
Analogous definitions apply also to $\Er$ and $\cHr=+2\Lambda\Er$.
We then have $E = \El + \Er$ being the total energy and $\cH = \cHl + \cHr$ the total helicity.
In the weakly nonlinear regime the amplitude equations can be written as
\begin{equation}
\deldelt{\hatL} = \fder{\cL}{\hatL} \quad\mbox{and}\quad
 \deldelt{\hatR} = \fder{\cL}{\hatR},
\label{eq:dlr} 
\end{equation}
where the simplest form of the Lagrangian is given by 
\begin{equation} \label{eq:lagrangian}
\cL[\hatL,\hatR] =  \gamma \left[|\hatL|^2 + |\hatR|^2 \right]
      -\mu \left[|\hatL|^4 + |\hatR|^4 \right] .
\end{equation}
The form of the Lagrangian is determined by the symmetry of the problem. 
The coefficient $\gamma$ is the linear growth rate and $\mu$ determines the
saturation of the instability in the weakly nonlinear regime. 
We emphasize that the $\mu$ and $\gamma$ for $\hatL$ and $\hatR$
could be different if and only if the chiral symmetry is broken
from the outset, but this is {\it not} the case here.
Now note that the Lagrangian must also be invariant under the parity transformation,
under which 
\begin{equation}
\cP(\hatL) = \hatR \quad{\rm and}\quad  \cP(\hatR) = \hatL.
\end{equation}
This additional symmetry allows one additional term in the Lagrangian given by
\begin{equation}
-\mus \left( |\hatL|^2|\hatR|^2 \right) .
\label{eq:mus} 
\end{equation}
With this additional term in the Lagrangian the evolution equations for the 
two eigenmodes are given by 
\begin{subequations}
\label{eq:evol}
\begin{align}
\deldelt{\hatL} &= \gamma \hatL -\left(\mu|\hatL|^2 + \mus|\hatR|^2\right)\hatL,
\label{eq:lt} \\
\deldelt{\hatR} &= \gamma \hatR -\left(\mu|\hatR|^2 + \mus|\hatL|^2\right)\hatR.
\label{eq:rt}
\end{align}
\end{subequations}
These equations, for certain parameters, allow and can describe the growth of one
handedness while the other is extinguished \cite{fau+dou+thu91}.
Similar equations, which describe the time dependence of the
amplitudes of the leading modes, but without considering their spatial
dependence, are often used to extend linear perturbation theory of 
hydrodynamic instabilities into the weakly nonlinear regime. 
In this form they are often called the Landau equations~\cite{lan-flu}. 

The energy of the left and right--handed modes is determined by, 
\begin{subequations}
\label{eq:dELR}
\begin{eqnarray}
\ddt{\El} &=& 2\gamma\El - 4\mu\El^2 - 4\mus\El\Er,\label{eq:dEl}\\
\ddt{\Er} &=& 2\gamma\Er - 4\mu\Er^2 - 4\mus\El\Er.\label{eq:dEr} 
\end{eqnarray}
\end{subequations}
These equations show that both $\El$ and $\Er$ grow exponentially
at the rate $2\gamma$ until nonlinear effects become important and
either $\El$ or $\Er$ saturates at $E_0\equiv\gamma/2\mu$ and the energy of
the mode of opposite handedness vanishes.
In principle, the achiral solution with
$\El=\Er\equiv E_{\rm a}=\gamma/2(\mu+\mus)$ is also possible, but,
as we will see below, such a solution is unstable for
$\mu<\mus$, which is what we find \Sec{DNS}.
The reason for this instability is the presence of the term
proportional to $\mus$, which represents
a phenomenon known as ``mutual antagonism'' in studies of the
origin of homochirality of bio-molecules \cite{Frank,Sandars,BAHN}.
We will return to this issue in \Sec{Homochirality}, where we discuss
the analogy with chiral symmetry breaking in biomolecules in more detail.

\begin{figure}
\includegraphics[width=180pt]{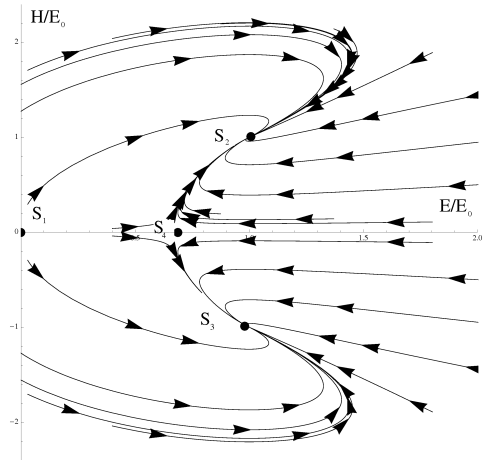}
\caption{The phase portrait for $\mu<\mu_*$. This is the 
typical situation in which $S_2$ and $S_3$ are attractive and $S_4$ is a saddle point.}
\label{fig:stabt}
\end{figure}

Using \Eqs{eq:eandh}{eq:evol} and defining $H=\cH/{2\Lambda}$
we have $E_{R}=(E + H)/2$ and $E_{L}=(E - H)/2$.
We can thus obtain the following evolution equations 
\begin{subequations}
\label{w}
\begin{eqnarray}
\ddt{E} &=& 2\gamma E - 2(\mu + \mus)E^2 -2(\mu - \mus){H^2}, \label{eq:let}\\
\ddt{H} &=& 2\gamma H - 4\mu E H. \label{eq:lht} 
\end{eqnarray}
\end{subequations}
The dynamical system described by (\ref{w}) and depicted in \Fig{fig:stabt}
has four fixed points in the $(E,H)$ plane, 
$S_1=(0,0)$, $S_{2,3} =(E_0,\pm E_0)$,  and  
$S_4=(2E_{\rm a}, 0)$ with eigenvalues $\lambda_1=(2 \gamma,2 \gamma)$, 
$\lambda_{2}=\lambda_3=(-2\gamma, 2(\mu-\mu_*)/\gamma)$, and 
$\lambda_4= (-2\gamma, 2\gamma - 4\gamma \mu / ( \mu+\mu_*) )$. 
The origin is always repulsive while $S_2$ and $S_3$ are sinks or saddle points
depending on the values of parameters $\mu$ and $\mu_*$.
$S_4$, corresponding to the achiral solution, can be an attractive point
only if $\mu_*<\mu$, otherwise is a saddle point.

A discussion of the amplitude equations is now in order. 
Firstly, we have assumed that there are exactly two modes of opposite
helicity that becomes critical at the onset of the instability.
This assumption is based on linear perturbation analysis.
As all the other modes in this case stable, in the spirit of
center manifold reduction, we have ignored their contributions to
total energy and helicity. 
If several modes are simultaneously unstable at the onset, then
we may expect a higher degree of complexity. 
Secondly, as our approach is based on symmetry, the form of the amplitude
equations that we obtain is very general. 
This is also a weakness of our approach, as we cannot determine the
expression for either $\mu$ or $\mu_*$. 
In principle, the method of multiscale expansion or center manifold reduction can be
applied to this problem to derive analytical expressions of $\mu$ and
$\mu_*$, but this is a difficult proposition in the present case as a
solution of the linear problem itself is not known in an analytically
closed form.

We can then compute the quantities $\gamma$, $\mu$ and $\mu_*$ with the help 
of direct numerical simulations (DNS)  by comparing the time evolution obtained for the left-hand side
of  (\ref{w}) in the weakly nonlinear phase where our description is valid.
We can anticipate that in most of our simulations $\mu<\mu_*$ and therefore 
the system should relax to a state of finite helicity in a finite time,
although we start from an infinitesimally small helicity. 
This is precisely what we observe in our DNS.

\section{Direct numerical simulations}
\label{DNS}

To analyze the evolution of the Tayler instability 
we choose our numerical domain to be a cylindrical shell
with an inner radius $s_{\rm in}=1$, outer radius $s_{\rm out}=3$,
and height $h=2$.
We perform simulations of the time-dependent resistive magnetohydrodynamic
equations for a compressible isothermal gas: the pressure is thus given by
$p=\rho\cs^2$, where $\rho$ is the density and $\cs$ is
the isothermal sound speed.

We use the \textsc{Pencil Code} \footnote{\texttt{http://pencil-code.googlecode.com/}}
to solve the equations for the magnetic vector potential $\AAA$,
($\BB = \curl \AAA$)
the velocity $\UU$, and the logarithmic density $\ln\rho$ in the form
\EQ
{\partial\AAA\over\partial t}=\UU\times\BB+\eta\nabla^2\AAA,
\label{dAdt}
\EN
\EQ
{\DD\UU\over\DD t}=-\cs^2\nab\ln\rho+\JJ\times\BB/\rho +\FF_{\rm visc},
\label{dUdt}
\EN
\EQ
{\DD\ln\rho\over\DD t}=-\nab\cdot\UU,
\label{dlnrhodt}
\EN
where
$$\FF_{\rm visc}=\rho^{-1}\nab\cdot2\nu\rho\SSSS$$ is the viscous force,
$\SSSS$ is the traceless rate of strain tensor having components
${\sf S}_{ij}=\half(U_{i,j}+U_{j,i})-\onethird\delta_{ij}\nab\cdot\UU$,
$$\JJ=\nab\times\BB/\mu_0$$ is the current density,
$\nu$ is the kinematic viscosity, 
and $\eta$ is the magnetic diffusivity.

We choose periodic boundary conditions in the vertical ($z$) and 
azimuthal ($\varphi$) directions, 
while at radial ($s$) boundaries we select perfectly conducting
boundary condition for the magnetic field and  stress-free boundary conditions
for velocity.
The resolution of the simulations presented here is
$128^3$ meshpoints in all three directions, but comparison with
different resolution demonstrated that our results are converged.

We choose a basic state with zero velocity and zero axial component
of the magnetic field ($B_z$).  
The azimuthal component of the magnetic field is 
\EQ
B_\varphi= B_0 \; (s/s_0) \exp [-(s-s_0)^2/\sigma^2],
\label{basic}
\EN
where $B_0$ is a normalization constant, $s_0=2$ and $\sigma=0.2$.
We choose $B_0$ and $\cs$ in such a way that the sound speed is much
larger than the Alfv\'en speed.
In this way we avoid magnetic perturbations to be dominant over
hydrodynamical perturbations.

In the basic state the Lorentz force due to the magnetic field is
balanced by the gradient of pressure.
Hence the pressure of the fluid is given by
\begin{eqnarray}
p&=&p_0-\frac{B_0^2}{4s_0^2}\,\Bigg[(2\,{s}^{2}-{\sigma}^{2}){{\rm e}^{-2\,{\frac { \left( s-{\it s_0} \right) ^{2}}{{
\sigma}^{2}}}}} \nonumber \\
&&\,+{\it s_0}\,\sigma\,\sqrt {\pi }\sqrt {2}
{{\rm erf}\left({\frac {\sqrt {2} \left( s-{\it s_0} \right) }{\sigma}}\right)}
 \Bigg],
\end{eqnarray}
where $p_0$ is a constant that must be large enough to ensure that the
pressure is positive.
If no perturbation is added, the system remains stationary.
Therefore, we add at the beginning of the simulation
a perturbation of the magnetic field with an infinitesimally small
net helicity given by the following expression:
\begin{equation}
\AAA=\delta s \cos \left(z \frac{n_z\pi}{h}\right)
\begin{pmatrix}
\sin m \varphi\cr0\cr \cos m \varphi
\end{pmatrix},
\end{equation}
where $\delta$ is an arbitrary small amplitudes which we set to $10^{-7}$ for all the simulations
and $k_z=q/s_{\rm in}=n_z\pi/h$ is the vertical wavenumber of the perturbation.

As discussed in \cite{bo08b}  kink instabilities are special 
case of the so called quasi-interchange instabilities, where combined
azimuthal and vertical field are present in the basic state.
In the incompressible limit the unstable eigenmodes 
can be described by a ($t,z,\varphi$)-dependence of the type
$\propto \exp (\gamma t  - i k_z z - i m \varphi)$
where the growth rate $\gamma$ is determined from a numerical
solution of the nonlinear eigenvalue problem 
for the radial disturbance $v_{1s}$ 

\begin{eqnarray}
&&\frac{d}{ds} \left[ \frac{1}{\lambda} (\gamma^2 + \omega_A^2) \left( 
\frac{d v_{1s}}{ds} + \frac{v_{1s}}{s} \right) \right] 
- k_z^2 (\gamma^2 + \omega_A^2) v_{1s}
\nonumber\\
&&+2 \omega_{B} \Big[ \frac{m (1 + \lambda)}{s^2
\lambda^2} \left( 1 - \frac{\beta\lambda}{1 + \lambda} \right) (\omega_{Az} +
2 m \omega_{B} ) \label{due}
\\
&&+ \frac{m \omega_{Az}}{s^2 \lambda^2} 
- k_z^2 \omega_{B} (1 - \beta)  \Big] v_{1s} + 
\frac{4 k_z^2 \omega_{A}^2 \omega_{B}^2}{\lambda (\gamma^2 + \omega_{A}^2)} v_{1s}
=0. \nonumber
\end{eqnarray} 
Here $\omega_{A}=( \BB \cdot \kk )/
\sqrt{\rho}$ with $\kk=(0,m/s,k_z)$, so
$\omega_{Az} =  k_z B_z/ \sqrt{ \rho}$.
Furthermore, we have defined $\omega_{B} = B_{\varphi}/s \sqrt{ \rho}$ and 
and $\lambda = 1 + m^2/s^2 k_z^2$.

Equation~(\ref{due}) describes the stability problem as a nonlinear eigenvalue problem.
This equation was first derived by Freidberg \cite{frei70} in his study of 
MHD stability of a diffuse screw pinch (see also \cite{bo08b}). The author found
that, for a given value of $k_z$, it is possible to obtain multiple values of the
eigenvalue $\gamma$, each one corresponding to a different eigenfunction, and 
calculated $\gamma$ for the fastest growing fundamental mode. The most general
form of Eq.~(\ref{due}), taking into account compressibility of plasma, was derived by 
Goedbloed \cite{goed71b}. Since we study the  stability assuming that the magnetic energy
is smaller than the thermal one, the incompressible form of Eq.~(\ref{due}) can be a sufficiently  
accurate approximation,  as we have verified. 
In our the case at hand,  $(\BB \cdot \kk)/\sqrt{\rho}=m\omega_B = \omega_A$ because we are interested in pure
kink (Tayler) instabilities, with $B_z\equiv 0$ in the basic state.
Note that as $\omega_{Az}=0$ in our case, Eq.(\ref{due}) is invariant for $m\rightarrow -m$.

In this latter case, once (\ref{due}) is solved and $v_{1s}$ is obtained,
the expressions for the other perturbed quantities
denoted by the subscript ``1" , read
\begin{subequations}
\label{pert}
\begin{eqnarray}
&&B_{1s}=-\frac{i}{\gamma s } B_\varphi \,v_{1s}, \\
&&B_{1\varphi}=-\frac{i}{\gamma s} B_\varphi  \,v_{1\varphi} - \frac{B_{\varphi}}{\gamma s}(\beta-1) v_{1s}, \\
&&B_{1z}=-\frac{i}{\gamma s} B_\varphi \, v_{1z}, \\
&&v_{1\varphi}=\frac{-i m }{(k_z s)^2 \lambda}  \frac{\partial}{\partial s} (s  \, v_{1s})-\frac{2 i m \omega_B^2 \, v_{1s}}{\lambda( \gamma^2+m^2 \omega_B^2)},\\
&&v_{1z} = -\frac{i}{k_z s} \frac{\partial}{\partial s} (s \, v_{1s})-\frac{m}{k_z s } v_{1\varphi}.
\end{eqnarray}
\end{subequations}
Unfortunately even for the case of pure kink instabilities, (\ref{due}) cannot be solved analytically and
one has to determine the dispersion relation numerically.
Therefore, to test our numerical setup we have solved numerically
Eq.~(\ref{due}) for the basic state
(\ref{basic}) for various values of $B_0$ and $\sigma$ in the limit
of small $v_A/c_s$ ratio to check that 
in the linear phase the growth rate extracted from the DNS is in agreement with the linear theory. 
In particular, as the inner radius of the cylinder is not at $s=0$,
we have set $v_{1s}=0$ at both inner and outer boundaries.
Note here that the growth rate and eigenfunctions of this instability
are known for the ideal MHD limit.
Hence to compare with those results, we choose viscosity and magnetic
diffusivity such that the dissipative time scales are much  larger
than the characteristic growth time (inverse of $\gamma$) of the
instability.

\begin{figure}
\includegraphics[width=\columnwidth]{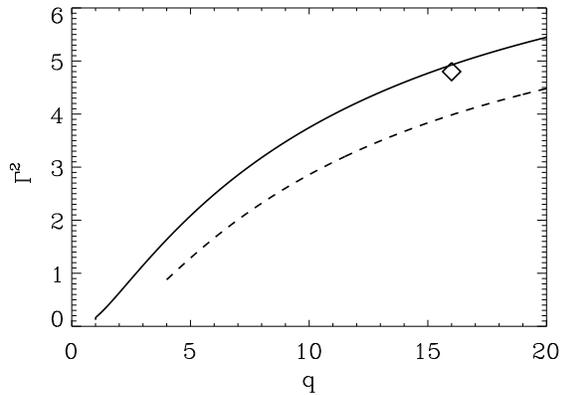}
\caption{
The dispersion relation for the dimensionless growth rate $\Gamma$ for the $m=\pm 1$ mode (solid line) and for the $m=\pm 2$ mode
(dashed). Higher values of $|m|$ have even smaller growth rates.
 This curve is obtained for a linear model with physical parameters corresponding to the nonlinear model {\tt Held},
 for which we indicate, with a rhombus,
the growth rate for its faster growing mode.
}
\label{fig:gr}
\end{figure}

The results are shown in Fig.~\ref{fig:gr} for the dimensionless growth
rate $\Gamma= \gamma t_A$, where
$t_A=s_{\rm out} \sqrt{\rho} / B_0$ is the Alfv\'en travel time,
as a function of the dimensionless vertical wave number
$q=k_z s_{\rm in}$ for model {\tt Held}  in Table~\ref{tab:models}, with $B_0=0.5$, and $n_z=10$.
In particular to compare the growth rate obtained from our DNS 
we have determined the characteristic vertical wave number of the unstable 
mode in the linear phase by means of the Fourier analysis of the magnetic fields.
We also found that in all the simulations the azimuthal wave number of the fastest growing mode
turned out to be always $m=\pm 1$ as higher values of $|m|$ have a smaller
growth rate as shown in Fig.~\ref{fig:gr}). 
We found that the corresponding growth rate determined
from the linear phase of our direct numerical simulation is
about $7-5\%$ smaller than the linear value, we think this
acceptable in view of unavoidable numerical diffusion in
three-dimensional numerical simulations.  

\begin{figure}
\includegraphics[width=.8\columnwidth]{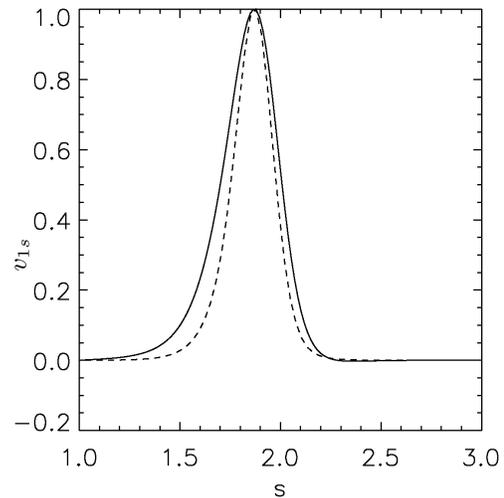}
\caption{
Eigenfunction  $v_{1s}$ for the $m=1$ mode for $q=16$. The result of the simulation, 
model {\tt Held} in Table~\ref{tab:models} (solid line) is overplotted
on the eigenfunction obtained solving \eq{due} (dashed line); see \cite{bo08b} for more details.
This is observed at $t/t_A=9$, that is during the linear growth of the instability.
}
\label{fig:eig}
\end{figure}

We see that the eigenfunction 
is rather localized for $q\gg 1$, as visible in the 
example shown in \Fig{fig:eig}.
We can exploit this property to obtain approximate explicit expressions
for the growth rate at large values of $q$.
In fact we can consider the magnetic field approximately constant around $s=s_0$ and apply the small-gap approximation  
(see \cite{bo08b} for details) so that  $v_{1s}\propto \sin (\pi (s-s_{0})/\sigma)$  and the dimensionless growth rate reads
\begin{eqnarray}
\Gamma^2&=&-\frac{2 c \Delta ^2 m^2 \left((\beta -1) m^2+(\beta -3) q^2\right)}{\left(m^2+q^2\right)
   \left(\Delta ^2 \left(m^2+q^2\right)+\pi ^2\right)} \\
   &&
 +  \frac{2 (\beta -1) \Delta ^2 q^2-c^2 m^2 \left(\Delta ^2 \left(m^2-3 q^2\right)+\pi^2\right)}{\Delta ^2 \left(m^2+q^2\right)+\pi ^2},
 \label{appr}
\nonumber
\end{eqnarray}
where $c=B_\phi /B_0 \approx \const$ and $\Delta = 2\sigma/s_0$.  
In the limit $q\gg1$,
despite the uncertain approximation that we have performed,
expression (\ref{appr}) differs by only $20\%$ from the numerical solution.

It is interesting to notice that, by using (\ref{pert}) in the limit $q\gg1$ we obtain the explicit expressions
\begin{eqnarray}
&&\langle \vv_1 \cdot \nabla \times \vv_1\rangle \approx -\frac{4 m \omega_B^2(\gamma^2 - m^2 \omega_B^2) \langle v_{1s}^2 \rangle}
{s_0^2 k_z (\gamma^2 +m ^2 \omega_B^2)^2},\\
&&\langle \BB_1 \cdot  \nabla \times \BB_1\rangle \approx \frac{4 m B_\phi^2 \omega_B^2 \langle v_{1s}^2 \rangle}
{s_0^4 k_z (\gamma^2 +m ^2 \omega_B^2)^2},
\end{eqnarray}
where the symbol $\langle \cdot \rangle$ denotes volume averaging. 
It is therefore clear that, at the nonlinear level, eigenfunctions
with non-zero $m$ will produce both kinetic and magnetic helicity 
whose sign will depend on the sign of $m$. 
The relevant point is that modes with opposite $m$ have identical growth rate, 
but opposite kinetic and magnetic helicity and the fate of the final helicity is
decided by the competition of modes with opposite azimuthal wavenumber. 

Moreover, according to the oscillation theorem \cite{gopo} as the
$m = \pm 1$ are unstable,
all the other modes with $m= \pm a$, where $a>1$ is a positive integer,  are also unstable,
but with a smaller growth rate. 
As a consequence, although in the linear phase the 
$m= \pm 1$ modes dominate the linear growth, already in the weakly
nonlinear phase the contribution of modes with 
$m\not= \pm 1$ can be also important for the selection of the final helical state.

\begin{figure}
\includegraphics[width=\columnwidth]{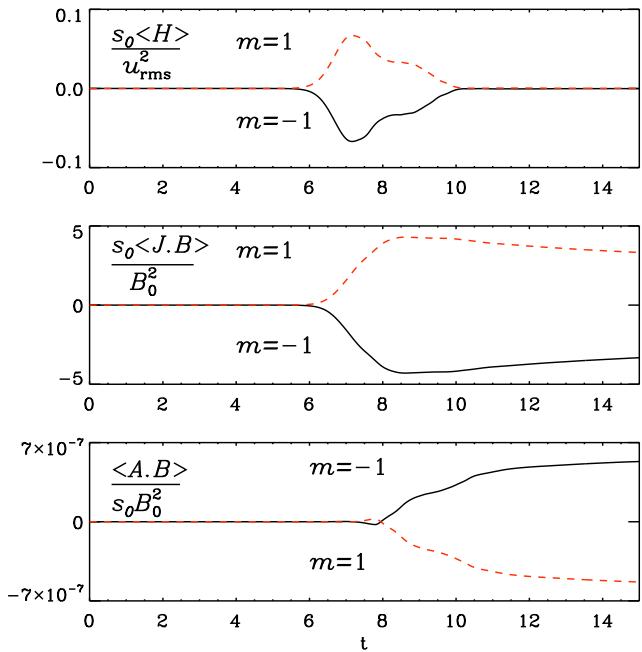}
\caption{(Color online) Kinetic, current, and magnetic helicities for two different runs 
(Models {\tt Hel} and {\tt Helm1} in Table~\ref{tab:models})
with helical perturbation and $m=\pm 1$.
$t$ is in units of the Alfv\'en travel time  $t_A$.
The viscous time is $t_{\nu} \approx 10^2*t_A$ and 
the magnetic diffusion is $t_{\eta} \approx 10^9*t_A$.
It is clear the difference in the evolutions of the kinetic, current and magnetic helicities, 
 in the first, second and third panel. 
 These plots show how these quantities grow with the same rate, but different sign,
 depending on the sign of the initial perturbation, that is the sign of $m$. 
 Note that for each model the magnetic helicity has opposite sign of the kinetic and current helicities.
}
\label{fig:comphel}
\end{figure}

The eigenfunctions appear clearly in our simulation
and they fit quite well the eigenfunctions calculated
by the linear model, as shown in \Fig{fig:eig}. 
In our simulations, during the growing phase of the instability
we observe a net increase of
the helicity, as shown in \Fig{fig:comphel} where we plot the 
time series of the normalized kinetic, current and magnetic helicity. 
It is interesting to notice that while kinetic helicity decays on 
the viscous time scale, the current and magnetic helicities 
reach a nonzero value at very large times.

In actual simulations we choose $\nu=10^{-2}$ (in code units), so that the viscous
time scale is $t_{\nu} = s^2/\nu \gg \gamma^{-1}$ and
the actual value of $\nu$ does not play a significant role in the weakly 
nonlinear phase as we verified in our simulations.
Moreover we decided to use a very small value for the magnetic diffusivity,
$\eta=10^{-9}$ (in code units).
 This is done to prevent the decay of the magnetic field by diffusion.
In general such small values of magnetic diffusivity would imply extremely
large values of magnetic Reynolds number which would be impossible to resolve with
the resolutions we use. 
Nevertheless we choose such values
to have a toroidal field stable on time scales
much longer than those of the instability.
However, in our simulations no sharp gradients of the magnetic field develop,
which is the reason why such small values of magnetic diffusivity are permissible.

\begin{figure}[t!]
\includegraphics[width=\columnwidth]{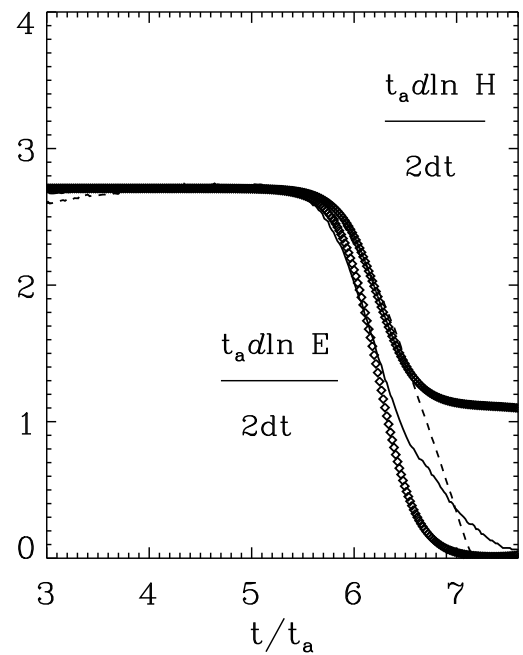}
\caption{ Time evolution for the logarithmic derivative of kinetic energy (solid line) $E$ and kinetic helicity $H$ (dashed) as measured in DNS
for models {\tt Hel} and {\tt Helm1}
(see \ref{tab:models}).
$t$ is in units of the Alfv\'en travel time  $t_A$.  
We overplot a fit of the model with equations \ref{w}. The best fit is obtained 
for $\gamma=2.71/t_A$, $\mu=7.5\cdot t_A/ s_{\rm out}^2$ and $\mus=18\cdot t_A/ s_{\rm out}^2$ 
and the solutions are over-plotted on the DNS results.
}
\label{fig:nonlin_hel}
\end{figure}

We can now determine the coefficients $\gamma$, $\mu$ and $\mu_*$ using
the time evolution of $H(t)$ and $E(t)$ obtained with our DNS
in solving the model (\ref{w}).
This can be done via a direct two-parameter $\chi^2$ minimization 
because the exponent $\gamma$ can easily be determined from the linear evolution
and one is left with only the determination of $\mu$ and $\mus$.
An example of this approach is depicted in Fig.~\ref{fig:nonlin_hel},
where the agreement with our numerical simulations 
is explicitly shown.
Note that around $t / t_A\approx 6.5$ we enter the deep nonlinear
phase and our treatment does not apply anymore.
We estimate this cutoff time for our simulations to be in the middle
of the decay transition for $d\ln H/dt$ and $d\ln E/dt$ depicted in
Fig.~\ref{fig:nonlin_hel} and we have checked that the values of
$\mu$ and $\mus$ are not strongly dependent on our cutoff time.

Our results are summarized in Table~\ref{tab:models}.
We see that the coefficients $\mu$ and $\mus$ are unchanged
for models that differ only in the sign of $m$ in the perturbation.
This is what we expect and one of the symmetries we have used 
to write the Lagrangian \eq{eq:lagrangian}.
Model {\tt Helc}  shows that the growth rate depends on the value of $c_s$,
but this does not change the values of $\mu$ and $\mus$.
Model {\tt Helb}  and model {\tt Held}  have a smaller growth rate due to a smaller $v_A/c_s$.
{\tt Helb}  has $\mu$ and $\mus$ smaller than {\tt Held} , due to the fact that in the latter
modes with higher $k_z$ have been excited by the initial perturbation.
Note that in our setup the ratio $v_A/c_s$ depends on $B_0$, but not on $c_s$.
This is due to the fact that the model is isothermal and the initial 
radial balance is obtained through a pressure gradient that balances the Lorentz force.
An increase of $n_z$ of the perturbation, as in model {\tt Heln10},
leads to a similar growth rate, but smaller $\mu$ and $\mus$. 
This can be explained saying that, while in the linear phase this model
evolves similarly to any $n_z=1$ model, in the weakly nonlinear phase the
evolution is different because of a faster growth of modes with higher $k_z$.
In our models we measure $2.2 \le \mus/\mu \le 2.6$ . 

\begin{table}
\caption{For every model $s$ goes from 1 to 3, $z$ from -1 to 1, 
the perturbation has an amplitude $\delta=10^{-7}$, $\sigma=0.2$. 
}
\vspace{12pt}\begin{tabular}{l|cccccccccc}
Model &$B_{0}^2/{p_0}$&$v_A/c_s$ &$c_s$&$m$ &$n_z$& $ \gamma \cdot t_A$ & $\mu\frac{ s_{\rm out}^2}{t_A}$ & $\mus\frac{ s_{\rm out}^2}{t_A}$ &$\mus/\mu$  \\
\hline
{\tt Hel}     &$10^{-1}$        &0.3     &10&$-1$&1         &2.71      &7.5     &18 &2.4\\
{\tt Helm1}   &$10^{-1}$        &0.3     &10&$+1$&1         &2.71      &7.5     &18 &2.4\\
{\tt Helc}    &$10^{-1}$        &0.3     &20&$-1$&1         &6.2       &7     &18.5 &2.6\\
{\tt Helb}    &$2.5\cdot10^{-2}$&0.15    &10&$-1$&1         &2.2       &1     &2.3  &2.3\\
{\tt Held}    &$2.5\cdot10^{-2}$&0.15    &10&$-1$&10        &2.2       &3     &7.3 &2.4\\
{\tt Heln10}  &$10^{-1}$        &0.3     &10&$-1$&10        &2.75      &4.5     &10 &2.2\\
\label{tab:models}\end{tabular}\end{table}

\section{Homochirality in biomolecules}
\label{Homochirality}

It is instructive to consider Eqs.~(\ref{eq:dELR}) as evolution equations for the
concentration of two molecules of opposite handedness, ${\sf L}$ and ${\sf R}$.
Let us assume that ${\sf L}$ and ${\sf R}$ can be synthesized from a substrate
${\sf S}$ through auto-catalytic reactions of the form
\begin{eqnarray}
{\sf S}\stackrel{\sf L}{\longrightarrow} {\sf L},\quad
{\sf S}\stackrel{\sf R}{\longrightarrow} {\sf R}.
\label{SLR}
\end{eqnarray}
Autocatalytic reactions of this type have been confirmed
in laboratory experiments \cite{Soai}.
Let us furthermore assume that ${\sf L}$ and ${\sf R}$ are capable of
polymerizing to form homochiral dimers,
\begin{eqnarray}
{\sf L}+{\sf L}\stackrel{\mu}{\longrightarrow} {\sf LL},\quad
{\sf R}+{\sf R}\stackrel{\mu}{\longrightarrow} {\sf RR},
\label{LLRR}
\end{eqnarray}
as well as heterochiral dimers,
\begin{eqnarray}
{\sf L}+{\sf R}\stackrel{\mus}{\longrightarrow} {\sf LR},
\label{LR}
\end{eqnarray}
then the evolution equations for the various concentrations are
\begin{subequations}
\label{eq:chemistry}
\begin{eqnarray}
\ddt{[{\sf S}]}&=&-k_C[{\sf S}]([{\sf L}]+[{\sf R}]),
\\
\ddt{[{\sf L}]}&=&k_C[{\sf S}][{\sf L}]-2k_S[{\sf L}]^2-2k_I[{\sf L}][{\sf R}],
\label{eq:chemistryL}
\\
\ddt{[{\sf R}]}&=&k_C[{\sf S}][{\sf R}]-2k_S[{\sf R}]^2-2k_I[{\sf L}][{\sf R}],
\label{eq:chemistryR}
\\
\ddt{[{\sf LL}]}&=&k_S[{\sf L}]^2,
\\
\ddt{[{\sf RR}]}&=&k_S[{\sf R}]^2,
\\
\ddt{[{\sf LR}]}&=&k_I[{\sf L}][{\sf R}],
\end{eqnarray}
\end{subequations}
which obeys the conservation law \cite{BAHN}
\begin{equation}
[{\sf S}]+[{\sf L}]+[{\sf R}]+2[{\sf LL}]+2[{\sf RR}]+2[{\sf LR}]=\const.
\end{equation}
These equations represent a subset of a more general polymerization
model \cite{Sandars}.
Comparing with \Sec{Amplitude} we see that \Eqs{eq:chemistryL}{eq:chemistryR}
are {\em identical} with \Eqs{eq:dEl}{eq:dEr} when substituting
$[{\sf L}]=\El$ and $[{\sf R}]=\Er$, and identifying
\begin{equation}
k_C=2\gamma,\quad
k_S=2\mu,\quad
k_I=2\mus.
\end{equation}
Hence we demonstrate quantitatively that the spontaneous production of helicity
from the fully nonlinear system of hydromagnetic equations can be described
by the simple model equations (\ref{eq:dELR}), which in turn represent
a simple set of chemical reactions \eqss{SLR}{LR}.

The analogy with homochirality in biochemistry is useful because it
helps identifying the phenomenon of mutual antagonism as the main cause
of chiral symmetry breaking.
This effect corresponds to a contribution to the nonlinear terms that
result from the interaction between modes of opposite handedness.
These are the terms proportional to $\mus$ and $k_I$ in
\Eqs{eq:dELR}{eq:chemistry}, respectively.
In the synthesis of polynucleotides this is known as enantiomeric
cross-inhibition and has been identified in laboratory experiments
\cite{Joyce}.
The synthesis of heterochiral dimers is essential in that it corresponds
to the production of waste needed to eliminate building blocks of that
handedness that is later is to disappear completely.

\section{Conclusions}
\label{Conclusion}

We have shown how net helicity is produced by the addition
of a small helical perturbation to a
non-helical system, thus driving the system to a final state characterized by
a finite value of the helicities and, therefore, breaking the initial symmetry.
We have shown further that this spontaneous symmetry breaking can be 
described by weakly nonlinear amplitude equations (\ref{w}). 
Furthermore, we have numerically determined the coefficients appearing in
the weakly nonlinear amplitude equations (\ref{w}) for the Tayler
instability.
Direct numerical simulations show that the ratio between the 
coefficients describing the weakly nonlinear phase is almost constant.
The agreement between the analytical model and the numerical solutions
is rather good in the beginning of the weakly nonlinear phase,
as shown in \Fig{fig:nonlin_hel}.
This demonstrates quantitatively the close analogy between helicity
production in hydromagnetic flows and the development of homochirality
in biochemistry, which is described by the same system of equations
as those resulting from the amplitude equations of the weakly nonlinear
model of the Tayler instability.
It will be useful to extend our analysis by means of a Landau-Ginzburg description 
of the amplitude equation by including non-homogeneous term in our Lagrangian to discuss
the possible pattern formation in this type of spontaneous chiral symmetry breaking.
We hope to address this issue in a following communication.

\section{Acknowledgements}
The authors thank P.\ Chatterjee and M.\ Rheinhardt for useful discussions.
A part of the work was performed when AB visited NORDITA under the program 
``Dynamo, Dynamical Systems and Topology''. 
FDS acknowledges HPC-EUROPA for financial support.
Financial support from European Research Council under the AstroDyn
Research Project 227952 is gratefully acknowledged.

\input{ref.tex}
\end{document}

%% file: symdef.inc
\newcommand{\BoldVec}[1]{\mathchoice%
  {\mbox{\boldmath $\displaystyle     #1$}}%
  {\mbox{\boldmath $\textstyle        #1$}}%
  {\mbox{\boldmath $\scriptstyle      #1$}}%
  {\mbox{\boldmath $\scriptscriptstyle#1$}}%
}
\newcommand{\EQ}{\begin{equation}}
\newcommand{\EN}{\end{equation}}
\newcommand{\EQA}{\begin{eqnarray}}
\newcommand{\ENA}{\end{eqnarray}}
\newcommand{\eq}[1]{(\ref{#1})}

\newcommand{\Eqs}[2]{Eqs.~(\ref{#1}) and~(\ref{#2})}

\newcommand{\eqss}[2]{(\ref{#1})--(\ref{#2})}
\newcommand{\Sec}[1]{Sect.~\ref{#1}}

\newcommand{\Fig}[1]{Fig.~\ref{#1}}

 \newcommand{\hatL}{\hat{{\bm L}}}
 \newcommand{\hatR}{\hat{{\bm R}}}

 \newcommand{\bx}{{\bm x}}

\newcommand{\xx}{\BoldVec{x}{}}

\newcommand{\vv}{\BoldVec{v} {}}

\newcommand{\UU}{\BoldVec{U} {}}

\newcommand{\BB}{\BoldVec{B} {}}

\newcommand{\AAA}{\BoldVec{A} {}}

\def \nn {\bm{n}}
\def \xx {\bm{x}}
\def \L    {\bm{L}}
\def \R    {\bm{R}}
\def \cH {\mathcal{H}}
\def \cHl {\mathcal{H}_L}
\def \cHr {\mathcal{H}_R}
\def \mus {\mu_{\ast}}
\def \El {E_L}
\def \Er {E_R}
\def \cL    {\mathcal{L}}
\def \cP    {\mathcal{P}}

\newcommand{\ddt}[1]{\frac{d #1}{dt}}
\newcommand{\deldelt}[1]{\frac{\partial #1}{\partial t}}
\newcommand{\fder}[2]{\frac{\delta #1}{\delta #2}}
\def \curl {{\bm \nabla}\times}

\newcommand{\JJ}{\BoldVec{J} {}}

\newcommand{\FF}{\BoldVec{F} {}}

\newcommand{\kk}{\BoldVec{k} {}}

\newcommand{\nab}{\BoldVec{\nabla} {}}

\newcommand{\SSSS}{\bm{\mathsf{S}}}

{}

\newcommand{\DD}{{\rm D} {}}

\newcommand{\const}{{\rm const}  {}}

\def\cs{c_{\rm s}}

\def\half{{\textstyle{1\over2}}}

\def\onethird{{\textstyle{1\over3}}}

\newcommand{\yana}[3]{, Astron. Astrophys. {\bf #2}, #3 (#1).}

\newcommand{\ymn}[3]{, Mon.\ Not.\ R.\ Astron.\ Soc.\ {\bf #2}, #3 (#1).}

\newcommand{\ynat}[3]{, Nature {\bf #2}, #3 (#1).}

\newcommand{\ypre}[3]{, Phys.\ Rev.\ E {\bf #2}, #3 (#1).}
\newcommand{\yprl}[3]{, Phys.\ Rev.\ Lett.\ {\bf #2}, #3 (#1).}

\newcommand{\yoleb}[3]{, Orig. Life Evol. Biosph. {\bf #2}, #3 (#1).}
\newcommand{\yapj}[3]{, Astrophys. J. {\bf #2}, #3 (#1).}

\newcommand{\yjour}[4]{, #2 {\bf #3}, #4 (#1).}

%% file: paper.bbl
\begin{thebibliography}{1}

\bibitem{ume}
H. Umezawa, {\it Thermo Field Dynamics and Condensed States}, Elsevier
(1982).

\bibitem{Gol}
N. D. Goldenfeld, {\it Lectures on Phase Transitions and the Renormalisation Group}, Addison-Wesley, (1992).

\bibitem{rb}
{\it Hydrodynamic Instabilities and the Transition to Turbulence}, 
edited by H.L. Swinney and J.P. Gollub (Springer-Verlag, New York, 1985), 2nd ed.;
M. C. Cross and P. C. Hohenberg, Rev. Mod. Phys. {\bf 65}, 851 (1993).

\bibitem{muse}
W. W. Mullins and R. F. Sekerka, J.App. Phys. {\bf 35}, 444 (1964).

\bibitem{exp}
E. Moses and V. Steinberg, Phys. Rev. A 34, 693 (1986);
A. J. Simon, J. Bechhoeffer, and A. Libchaber, Phys. 
Rev. Lett. 61, 2574 (1988); G. Faivre, S. de Cheveigne, 
C. Guthmann, and P. Kurowski, Europhys. Lett. 9, 779 
(1989); F. Melo and P. Oswald, Phys. Rev. Lett. 64, 1381 
(1990); H. Z. Cummins, L. Fourtune, and M. Rabaud, 
Phys. Rev. E 47, 1727 (1993).

\bibitem{seli93}
J. V. Selinger, Z.-G. Wang, R. F. Bruinsma, and C. M. Knobler, Phys. Rev. Lett. 70,
1139 (1993).

\bibitem{pinter06}
A. Pinter, M. L\"ucke, and C. Hoffmann\yprl{2006}{96}{044506}

\bibitem{cha+mit+bra+rhe11}
P. {Chatterjee}, D. {Mitra}, A. {Brandenburg}, and M. {Rheinhardt}\ypre{2011}{84}{025403R}
P. {Chatterjee}, D. {Mitra}, M. {Rheinhardt}, and  A. {Brandenburg}
\yana{2011}{534}{A46}

\bibitem{gel+rud+hol11}
M. Gellert, G. R\"udiger, and R. Hollerbach\ymn{2011}{414}{2696}

\bibitem[\protect\citeauthoryear{Tayler}{1973a}]{tay73a}
Tayler, R. J.\ymn{1973}{161}{365}

\bibitem[\protect\citeauthoryear{Tayler}{1973b}]{tay73b}
Markey, P., Tayler, R. J.\ymn{1973}{163}{77}




\bibitem[\protect\citeauthoryear{Bonanno \& Urpin}{2008a}]{bo08a}
A. Bonanno and V. Urpin\yana{2008}{477}{35}

\bibitem[\protect\citeauthoryear{Bonanno \& Urpin}{2008b}]{bo08b}
A. Bonanno and V. Urpin\yana{2008}{488}{1}


\bibitem[\protect\citeauthoryear{Bonanno \& Urpin}{2011}]{bu11}
A. Bonanno and V. Urpin\ypre{2011}{84}{056310}{}


\bibitem[\protect\citeauthoryear{Bonanno \& Urpin}{2012}]{bu12}
A. Bonanno and V. Urpin\yapj{2012}{747}{137}

\bibitem[\protect\citeauthoryear{Braithwaite \& Nordlund}{2006}]{brano06}
J. Braithwaite and \AA. Nordlund\yana{2006}{450}{1077}

\bibitem[\protect\citeauthoryear{Braithwaite}{2006}]{bra06}
J. Braithwaite\yana{2006}{453}{687}

\bibitem[\protect\citeauthoryear{Spruit}{1999}]{spru99}
H. Spruit\yana{1999}{349}{189}

\bibitem{fau+dou+thu91}
S. {Fauve}, S. {Douady}, and O. {Thual}, J. Phys. II {\bf 1},  311  (1991).


\bibitem[\protect\citeauthoryear{Qingzeng}{1997}]{qing}
F. Quingzeng, Applied Mathematics and Mechanics, {\bf 18}, {865} (1997).

\bibitem{lan-flu}
L.D. Landau and E.M. Lifshitz, {\it Fluid Mechanics, Volume 6 (Course of
  theoretical physics)}, Pergamon Press, 2nd English Ed., Translated from 
  {\it Gidrodinamika}, 3rd edition, "Nauka", Moscow, 1986, 
  Chap. 3 (1987).

\bibitem{Frank}
F. C. Frank\yjour{1953}{Biochim.\ Biophys.\ Acta}{11}{459}

\bibitem{Sandars}
P. G. H. Sandars\yoleb{2003}{33}{575}

\bibitem{BAHN}
A. Brandenburg, A. C. Andersen, S. H\"ofner, and M. Nilsson\yoleb{2005}{35}{225}

\bibitem{frei70}
J. Freidberg, Phys. Fluids. 13, 1812 (1970).

\bibitem{goed71b}
J. P. Goedbloed, Physica. 53. 535 (1971).

\bibitem[\protect\citeauthoryear{Goedbloed \eal}{2004}]{gopo}
J. P. H. Goedbloed and S. Poedts, Principles of Magnetohydrodynamics, CUP, (2004).

\bibitem{Soai}
K. Soai, T. Shibata, H. Morioka, and K. Choji\ynat{1995}{378}{767}

\bibitem{Joyce}
G. F. Joyce, G. M. Visser, C. A. A. van Boeckel, J. H. van Boom, L. E. Orgel, and J. Westrenen\ynat{1984}{310}{602}


\end{thebibliography}
